\documentclass[letterpaper]{article} 
\usepackage[]{aaai25}  
\usepackage{times}  
\usepackage{helvet}  
\usepackage{courier}  
\usepackage[hyphens]{url}  
\usepackage{amsmath}
\usepackage{graphicx} 
\usepackage{natbib}  
\usepackage{caption} 
\usepackage{algorithm}
\usepackage{algorithmic}

\usepackage{newfloat}
\usepackage{listings}
\DeclareCaptionStyle{ruled}{labelfont=normalfont,labelsep=colon,strut=off} 
\floatstyle{ruled}
\newfloat{listing}{tb}{lst}{}
\floatname{listing}{Listing}
\title{From Millions of Tweets to Actionable Insights: \\ Leveraging LLMs for User Profiling}
\author{
    Vahid Rahimzadeh\textsuperscript{\rm 1},
    Ali Hamzehpour\textsuperscript{\rm 1},
    Azadeh Shakery\textsuperscript{\rm 1, \rm 2},
    Masoud Asadpour\textsuperscript{\rm 1}
}

\affiliations {
    \textsuperscript{\rm 1}School of Electrical and Computer Engineering, College of Engineering, University of Tehran, Tehran, Iran
\\
    \textsuperscript{\rm 2}School of Computer Science, Institute for Research in Fundamental Sciences (IPM), Tehran, Iran
\\
    \{rahimzade,ali.hamzehpour,shakery,asadpour\}@ut.ac.ir
}

\usepackage{bibentry}

\begin{document}

\maketitle

\begin{abstract}
Social media user profiling through content analysis is crucial for tasks like misinformation detection, engagement prediction, hate speech monitoring, and user behavior modeling. However, existing profiling techniques, including tweet summarization, attribute-based profiling, and latent representation learning, face significant limitations: they often lack transferability, produce non-interpretable features, require large labeled datasets, or rely on rigid predefined categories that limit adaptability.

We introduce a novel large language model (LLM)-based approach that leverages domain-defining statements—key characteristics defining the important pillars of a domain as foundations for profiling. Our two-stage method first employs semi-supervised filtering with a domain-specific knowledge base, then generates both abstractive (synthesized descriptions) and extractive (representative tweet selections) user profiles. By harnessing LLMs' inherent knowledge with minimal human validation, our approach is adaptable across domains while reducing the need for large labeled datasets.

Our method generates interpretable natural language user profiles, condensing extensive user data into a scale that unlocks LLMs' reasoning and knowledge capabilities for downstream social network tasks. We contribute a Persian political Twitter (X) dataset and an LLM-based evaluation framework with human validation. Experimental results show our method significantly outperforms state-of-the-art LLM-based and traditional methods by 9.8\%, demonstrating its effectiveness in creating flexible, adaptable, and interpretable user profiles.

\end{abstract}

%
\section{Introduction}

User profiling on social media platforms has become increasingly crucial in the digital age, offering valuable insights for a wide range of applications including engagement prediction \cite{user_profile_egagement}, expert finding \cite{rostami2023deep}, hate speech detection \cite{profile_hate_importance}, fake news identification \cite{user_profile_fake_2,user_profile_fake}, and user behavior modeling \cite{user_profile_behavior,user_behavior_agents_s3}.
Previous approaches to user profiling have faced various limitations. Tweet summarization methods often rely on pre-existing domain-specific ontologies \cite{profile_summ_ontology} and complex architectures or large labeled datasets \cite{tssubert}. Attribute-centric methods typically depend on predefined categories, limiting their adaptability \cite{brazilian_presidential,attribute_centric_profile}. Latent representation learning methods, while effective for specific tasks \cite{task_specific_representation,task_specific_representation_2}, struggle with transferability across different applications. Despite efforts to address transferability and adaptability challenges \cite{social_llm,profile_embedding_adaptable}, existing approaches lack interpretability and require substantial labeled data. These limitations highlight a need for an interpretable, adaptable, and extendable user profiling method based solely on users' past content.

Large language models have transformed the field of Natural Language Processing (NLP) significantly in recent years and have demonstrated impressive performance in summarization tasks \cite{summarization_is_dead,human_prefered_llm_sum}. Despite this potential, incorporating LLMs' reasoning and summarization capabilities for social network analysis—particularly user profiling—remains heavily under-studied. 

Our approach addresses these challenges through a two-stage process. First, we employ a semi-supervised semantic filtering method to efficiently process large-scale social media data, identifying domain-relevant content while minimizing human annotation requirements. Building on this filtered data, our method leverages domain-defining statements that emerge directly from the data itself to generate both abstractive and extractive user profiles. Abstractive profiles synthesize a user's content history into concise natural language descriptions, enabling inference of deeper patterns and trends, while extractive profiles identify and select the most representative original posts, preserving the user's authentic voice. These complementary approaches create interpretable representations that capture key user characteristics within the target domain. Our LLM-based evaluation framework assesses profile quality by comparing performance on an open-book QA task using domain-defining statements as queries, with either the generated profiles or the user's full history serving as context, validating results against human-annotated ground truth.

Our comprehensive pipeline—encompassing data filtering, profile generation, and evaluation—leverages LLMs' inherent reasoning capabilities and general knowledge with minimal human validation, significantly reducing the need for large domain-specific labeled datasets while producing profiles in a natural language format readily usable by LLMs for downstream tasks.

Our work contributes: (1) a semi-supervised domain-specific filtering method for processing millions of tweets with minimal human annotation; (2) novel LLM-based abstractive and extractive profiling techniques producing interpretable natural language user profiles; (3) an LLM-assisted evaluation framework with human validation independent of downstream tasks; (4) a Persian political Twitter (X) dataset to be used for our profiling tasks; and (5) comprehensive analysis showing our approach outperforms existing methods by 9.8\%.

These contributions pave the way for more effective and flexible user profiling in social media analysis, leveraging the vast knowledge of LLMs to reduce dependency on labeled data while maintaining interpretability and adaptability across various domains.

The remainder of this paper covers: related work (Section 2); our novel semi-supervised filtering method for creating a Persian political Twitter (X) dataset (Section 3); LLM-based profiling methods and evaluation framework (Section 4); experimental setup (Section 5); and results analysis with discussion (Section 6).

\section{Related Works}

\subsection{Tweet Summarization}
Traditional tweet summarization has advanced from extractive models to Deep Neural Networks.~\cite{tweet_sum_deep_learning} proposed a two-stage framework with event detection via clustering and summarization using pre-trained deep models, achieving strong ROUGE scores.~\cite{tssubert} introduced a large dataset for 
er) event summarization and an extractive model combining pre-trained language models with frequency-based representations.~\cite{cheng-lapata-2016-neural} presented a neural framework for extractive summarization using a hierarchical encoder and attention-based extractor, achieving competitive results.
Recent research has explored LLMs for text summarization, including tweets.~\cite{llm_for_sum} compared four LLMs for abstractive summarization using various metrics.~\cite{tssubert} introduced a large dataset and a neural model combining pre-trained language models with vocabulary-based representations for tweet summarization.~\cite{mishra2023llm} demonstrated a semi-supervised approach for extractive dialog summarization using LLMs to generate pseudo-labels, achieving strong results on the TWEETSUMM dataset with limited labeled data. ~\cite{llm_new_sum} found that instruction tuning, not model size, is key for zero-shot summarization, with LLM-generated summaries being comparable to human-written ones despite stylistic differences.



\subsection{User Profiling}
User profiling has advanced from early stereotype models to sophisticated Machine Learning  approaches. Initially, methods like Case-Based Reasoning and Bayesian Networks built incremental profiles ~\cite{user_profiling_baysian}. Recent techniques involve deep learning, multi-behavior modeling, and graph data structures, constructing accurate profiles from rich interaction data ~\cite{Purificato2024UserMA}. These advancements enhance personalization in areas like information retrieval, recommender systems, and education ~\cite{Purificato2024UserMA}.~\cite{Kanoje2015UserPT}), with a shift towards implicit data collection and privacy-preserving techniques. Ongoing research focuses on improving accuracy, efficiency, and ethical considerations in AI-driven systems.
\newline
LLMs are transforming user profiling and personalization by reasoning through user activities and describing interests with nuance, leading to deeper insights and improved personalization ~\cite{christakopoulou2023large}. Integrated with text and graph-based methods, LLMs enhance user modeling applications~\cite{user_modeling_in_era}. The LaMP benchmark evaluates LLMs in generating personalized outputs across diverse language tasks~\cite{salemi-etal-2024-lamp}. LLMs shift personalization from passive information filtering to active user engagement, providing services through general-purpose interfaces~\cite{christakopoulou2023large}. Their ability to explore user requests proactively and deliver explainable information addresses personalization challenges, offering significant opportunities for advancing user profiling techniques.
\newline
We build on recent advancements by integrating LLMs with a semi-supervised filtering approach to improve user profiling on social media. Instead of relying on large labeled datasets and predefined categories, our method leverages LLMs’ reasoning and summarization abilities to generate interpretable profiles in natural language. This approach enhances adaptability across domains while reducing the need for extensive labeled data, addressing key limitations of current profiling techniques.
\begin{figure*}[!ht]
	\centering
	\includegraphics[width=0.9\linewidth]{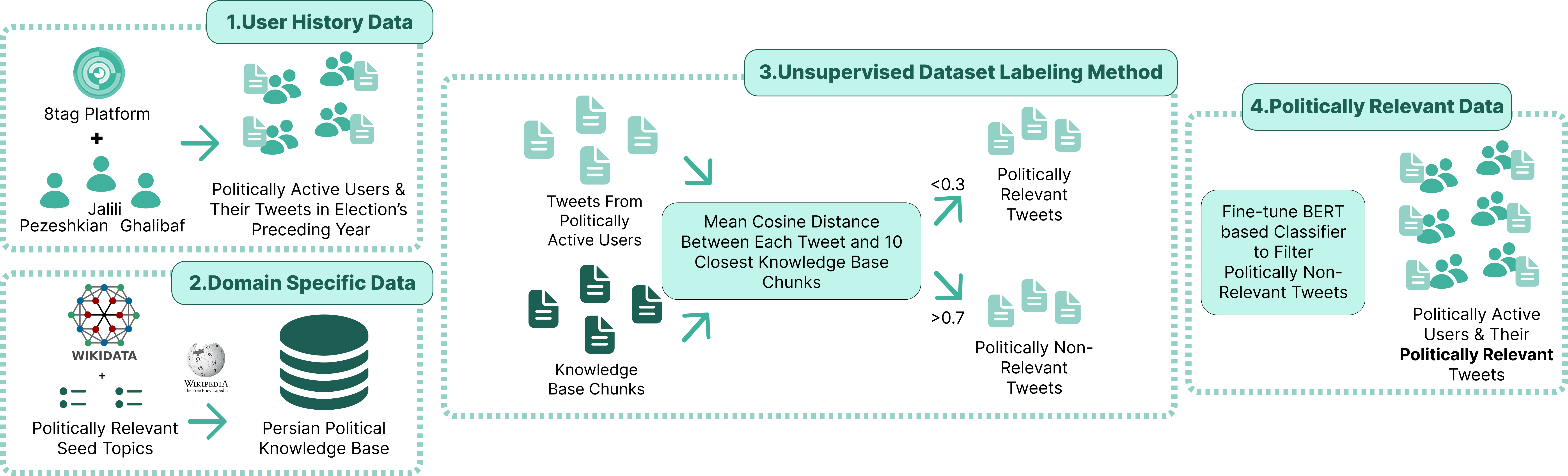}
	\caption{Dataset curation pipeline for political content related to the 2024 Iranian presidential election, including data collection, knowledge base construction, semi-supervised labeling, and BERT-based classification}
	\label{fig:dataset_curation_illustration}
\end{figure*}
\section{Dataset}

This section outlines how we created PersianPol6M (PP6M), a comprehensive dataset of political content related to the 2024 Iranian presidential election, using the pipeline illustrated in Figure~\ref{fig:dataset_curation_illustration}. We developed this dataset primarily to implement and evaluate our user profiling methods for social networks, though it also serves as a valuable resource for analyzing political discourse patterns and user stance\footnote{Please contact the authors for dataset access.}.

Our process begins with data collection (Step 1 in Figure~\ref{fig:dataset_curation_illustration}), where we gather over 6 million tweets from politically active users involved in the election campaign, along with their retweet interaction graph from 8tag\footnote{\url{https://www.8tag.ir}}. Since these users produce both political and non-political content, we need a scalable filtering algorithm to extract domain-specific tweets. We address this by constructing a domain-specific knowledge base (Step 2) focused on Persian politics, followed by implementing a novel semi-supervised filtering approach (Step 3) that leverages this knowledge base to identify and retain domain-relevant political content. Finally, we compile the filtered dataset (Step 4), resulting in PersianPol6M (PP6M) - a refined collection of tweets focused on Persian political content from politically engaged users.

While this curation process focuses on the political domain, our proposed method can be used for other domains by leveraging pre-existing domain-specific knowledge bases when available or by constructing new knowledge bases from structured resources such as Wikidata~\cite{wikidata}. Each step of this pipeline is elaborated in the corresponding subsections below, following the step numbers shown in Figure~\ref{fig:dataset_curation_illustration}.

\subsection{Dataset Collection}
\begin{figure}[ht]
	\centering
	\includegraphics[width=0.8\linewidth]{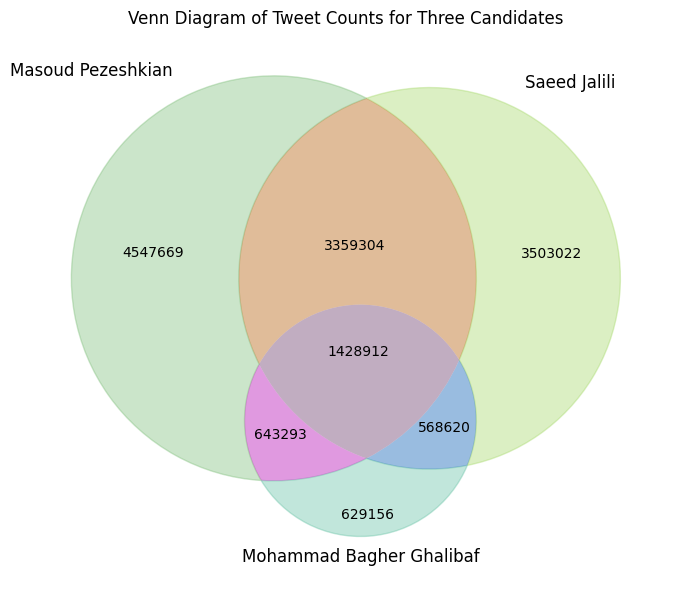}
	\caption{Venn diagram illustrating the overlap of unique tweets among the three main candidates in the 2024 Iranian Presidential Election: Pezeshkian, Jalili and Ghalibaf}
	\label{fig:venn_diagram}
\end{figure}

The initial dataset includes over 6 million tweets from the top 3000 users (1000 per candidate) who mentioned the names of the three primary candidates in the 2024 Persian Presidential Election: Pezeshkian, Jalili, and Ghalibaf. Collected from March 2023 to March 2024, this dataset provides a comprehensive view of the political discourse surrounding these candidates during this period. Furthermore, the Venn diagram (Figure~\ref{fig:venn_diagram}) illustrates a good balance between the two major political branches in the country, with Pezeshkian representing the Reformists and both Ghalibaf and Jalili representing the Principlists.
A retweet graph is also collected, which is used for community detection to obtain a representative data sub-sample for further analysis in subsequent sections.
\subsection{Domain-Specific Knowledge Base}
To identify and retain only political tweets from a user's tweet history without using labeled data, we propose a semi-supervised approach. We first build a Persian political knowledge base and then calculate similarity scores between each tweet and this knowledge base. These scores determine the political relevance of the tweets, enabling us to filter out non-political ones systematically.
\begin{figure*}
	\centering
	\includegraphics[width=1\linewidth]{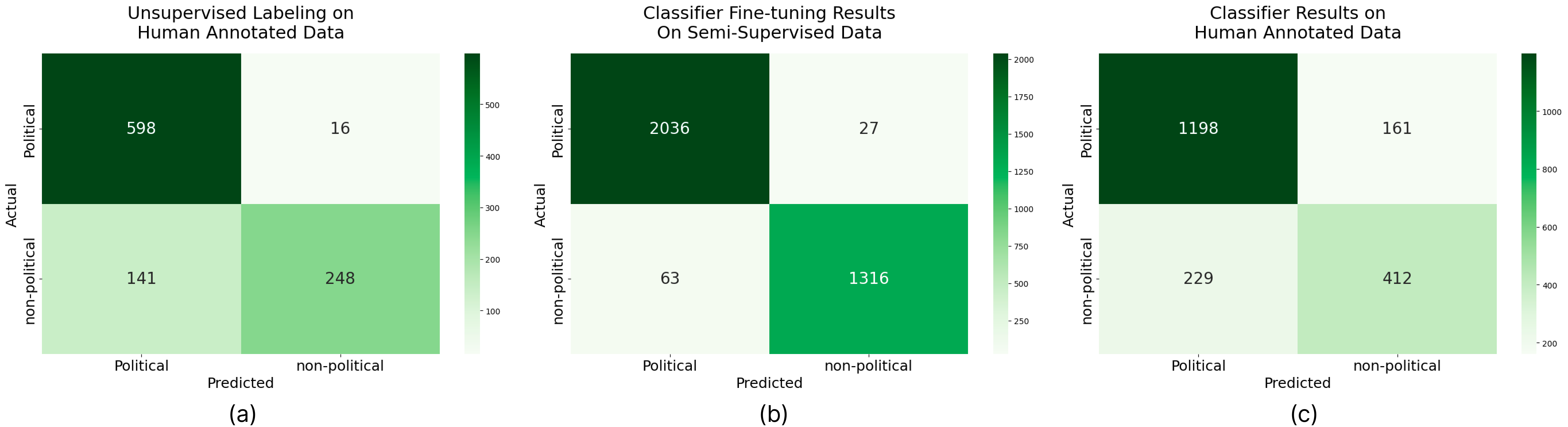}
	\caption{Confusion matrices for: (a) unsupervised labeling on human annotations, (b) classifier fine-tuning with semi-supervised data on test set, and (c) final classifier evaluation using borderline data annotated by human}
	\label{fig:confusion_matrix_classifier}
\end{figure*}

To construct the knowledge base, we follow a multi-step process: 1) Collect a set of seed Wikidata \cite{wikidata} entities, using examples such as elections held after the 1979 revolution, 2) Expand these entities by traversing Wikidata to a depth of three using specific edge types, such as "Candidate" and "Position held", 3) Include entities linked from the IRI opposition Wikipedia page to cover Iranian government opposition, and 4) Extract Persian Wikipedia content for each entity. This process results in a knowledge base with 698 nodes, 814 edges, and 731 documents.

\subsection{Filtering Domain-Specific User Tweets}
\subsubsection{Classifier Dataset Curation Process.}
We chunk the knowledge base using a markdown-aware method and compute embeddings for all tweets and chunks using the BGE-M3 \cite{bge-m3} model. Then, a cosine distance (1 - cosine\_similarity) is calculated between each tweet's embedding and all knowledge base chunk embeddings.
We determine the \textit{k} closest knowledge base chunks for each tweet based on distance values.
\begin{figure}
	\centering
	\includegraphics[width=0.99\linewidth]{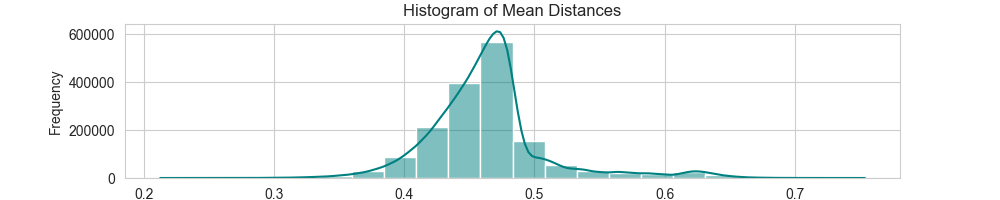}
	\caption{Histogram of mean distances between tweets and their top 10 closest knowledge base chunks. Lower  distances indicate stronger alignment with political content}
	\label{fig:similarity_distribiution}
\end{figure}
Tweets are labeled using the following criteria:
\begin{equation}
	\label{eq:tweet_labeling}
	\begin{cases} 
		1, & \text{if } \frac{1}{k} \sum_{i=1}^{k} d(T, C_i) < 1 - \theta \\
		0, & \text{if } \frac{1}{k} \sum_{i=1}^{k} d(T, C_i) > \theta
	\end{cases}    
\end{equation}
where \(T\) is the tweet embedding, \(C_i\) is the \(i\)-th closest chunk embedding, and \(k\) represents the number of closest chunks considered. In this context, 1 indicates a political tweet and 0 indicates a non-political tweet. Our empirical parameter tuning shows that setting \(\theta\) = \(0.7\)  and \(k\) = \(10\) yields high-quality labels (Figure~\ref{fig:similarity_distribiution} shows the distribution of mean distances between tweets and their closest knowledge base chunks).

This process labels 15,000 tweets (28\% political, 72\% non-political). 
We manually label 1,000 tweets to evaluate the accuracy of our semi-supervised approach (Figure~\ref{fig:confusion_matrix_classifier}a). The results show a high correlation with human annotation, achieving a precision of 93\% on positive political data.

\subsubsection{Political Tweets Classifier.}
We fine-tune the last three layers of a state-of-the-art Persian encoder, TookaBERT-Base \cite{tookabert} using our labeled dataset generated from the previous sections. The result is an efficient and flexible classifier. The performance of the classifier on the test set is illustrated in Figure~\ref{fig:confusion_matrix_classifier}b, achieving a macro f1-score of 97\%. Furthermore, we manually label 2000 borderline samples with similarity scores between \(\theta\) and 1-\(\theta\), as these cannot be labeled by our distance based approach confidently (See Equation~\ref{eq:tweet_labeling}). Resulting classifier achieves the same performance on these challenging samples as the distance based approach does on simple samples, while significantly reducing computational costs (see Figure~\ref{fig:confusion_matrix_classifier}c and Figure~\ref{fig:confusion_matrix_classifier}a).

Additionally, the classifier's training dataset can be improved by incorporating tweets related to other user profiling aspects, such as cultural or personality traits. For instance, if the goal is to identify political or aggressive content, the dataset can be augmented with aggressively toned tweets as either positive or negative examples. Furthermore, by leveraging multiple knowledge bases and targeted data points from diverse domains at the start, the classifier can be trained to filter content from various perspectives, resulting in greater comprehensiveness and efficiency.
\subsection{PersianPol6M}
The classifier is applied to the entire dataset, filtering tweets for political relevance. This reduces the dataset to over 1.7 million political tweets (Table~\ref{tab:summary_stats}), focusing on the 2024 Iranian Presidential Election candidates and broader political discourse. This refined dataset provides a valuable resource for studying political engagement and sentiment in Iran's digital landscape.
\captionsetup[table]{skip=2pt} 
\begin{table}[htbp]
	\centering
	
	\small 
	\begin{tabular}{l@{\hspace{0.5em}}ccc} 
		& Pezeshkian (\%) & Jalili (\%) & Ghalibaf (\%) \\
		\hline
		Total Twts & -70.18 & -80.18 & -73.17 \\
		Total Users & -0.51 & -34.47 & -0.54 \\
		Avg Twt/User & -70.03 & -69.76 & -73.02 \\
		Avg Twt Len & 70.82 & 86.83 & 81.31 \\
		Median Twt Len & 151.72 & 194.37 & 189.19 \\
	\end{tabular}
	\caption{Percentage changes in the total tweets, users, retweets, likes, and tweet lengths after filtering for each candidate}
	\label{tab:summary_stats}
\end{table}
\begin{figure*}
	\centering
	\includegraphics[width=\textwidth]{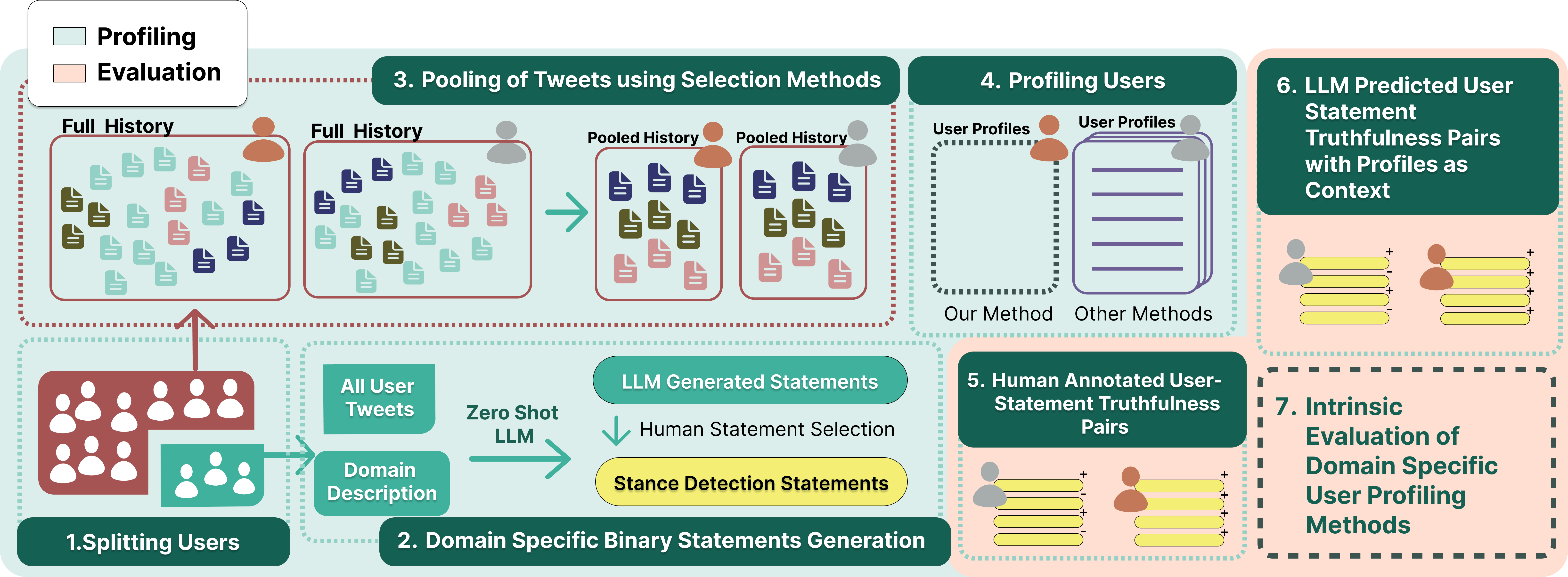}
	\caption{Step-by-step illustration of the user profiling and evaluation process. The pipeline includes (1) user splitting, (2) domain specific defining statement generation, (3) heuristic tweet selection to get a pool of tweets for annotation and profiling, (4) user profiling with tweet pool and statements (5) statement-user pairs ground truth generation by human using pool of tweets, (6) LLM-based evaluation on user-statement pairs with user profiles as context, and (7) profile comparison based answers with human ground truth 
	}
	\label{fig:method}
\end{figure*}
\section{Methodology}
To enable robust analysis, we begin by sampling representative users. Our methodology, illustrated in Figure~\ref{fig:method}, consists of two main components: user profiling (light green background, Steps 1-4) and evaluation (light orange background, Steps 5-7). After explaining our data sampling approach in Section 4.1, we detail our novel LLM-based profiling method in Section 4.2, followed by our intrinsic evaluation framework for measuring information preservation in Section 4.3.
\subsection{Data Sampling}
Employing Louvain \cite{modularity}, a widely used modularity based community detection algorithm, we perform community detection on PP6M. We then randomly select 10\% of users from the top 20\% communities based on user counts. This process yields a final sample of 150 users with approximately 140,000 tweets. The random seed is set to \(123\) in all steps. 

\subsection{LLM-Based User Profiling}
Our profiling method focuses solely on users' generated content history, with an emphasis on political domain profiles for this work. To ensure unbiased profiling, we first divide our dataset into two splits: one for determining domain \textit{defining statements} and another for profile generation. Our profiles are based on \textit{defining statements} that characterize the domain. These statements are extracted directly from the data, rather than being predefined or sourced from external knowledge. Given that stance detection on Twitter (X) data can effectively identify user political alignment ~\cite{importance_of_stance_in_user_profiling}, for this specific study, we implement our defining statements as \textit{stance claims}. We generate these stance claims from the first split and use them to profile users in the second split. For example, one such claim might be \textit{The user supports the foreign policies of Iranian government}. However, this approach is adaptable, allowing researchers to tailor the defining statements to their specific domain of interest while maintaining the core methodology (see Appendix~\ref{appendix:adaptation}). 

In the following subsections, we detail each step of our profiling methodology: first explaining how we split the population to avoid data contamination, then describing the process of generating defining statements, followed by a pooling step that makes human test set annotation feasible, and finally outlining how we create user profiles based on these statements.
\subsubsection{Population Splitting.}
To facilitate an unbiased profiling process, we divide our dataset into two distinct sets: 50 users for generating defining statements and 100 users for profiling. The splitting is done using stratified sampling from k-means clusters \cite{KMeans}, ensuring representativeness across different political orientations (Implementation details are available in Appendix~\ref{appendix:implementation-details}).

\subsubsection{Gathering Domain Defining Statements.}
Through zero-shot prompts, we generated more than 500 domain-descriptive stance claims, refined through deduplication and human expert selection, to identify 15 key Persian stance claims capturing the political landscape of Persian Twitter (X). These claims will act as the foundation aspects of our user profiling and evaluation methods. A translated statement from Persian is provided below:
\begin{itemize}
	\item \textbf{Example Statement: }The user's opinion about the country's economic situation is positive.
\end{itemize}
More translated examples of final claims, alongside the prompt and settings are available in the Appendix~\ref{appendix:data-details}.
This approach makes it possible to compare profiles between users/communities, unlike pure summarization methods, and enables precise evaluation of profile qualities through our proposed evaluation method in Section 4.3.
\subsubsection{Pooling on Tweets using Selection Methods.}
To manage the computational demands of LLM-based experiments across various settings and ensure feasible human annotations, we employed four tweet pooling methods to reduce the number of tweets per user: random selection, selecting the top 20 tweets nearest to the mean of user tweet embeddings, stratified sampling from k-means clusters on user tweet embeddings, and iterative elimination based on similarity. For the iterative elimination method, we start with the tweet closest to the mean embedding, then iteratively remove tweets that exceed a certain similarity threshold with the selected tweet, repeating this process until we have 20 tweets. This process yields 80 relevant tweets per user-statement pair (20 from each method), which will serve as the user's history for subsequent profiling tasks. The relevant algorithms are available in Appendix~\ref{appendix:implementation-details}. 
\subsubsection{User Profiling.}
\begin{figure}[!tp]
	\centering
	\includegraphics[width=1\linewidth]{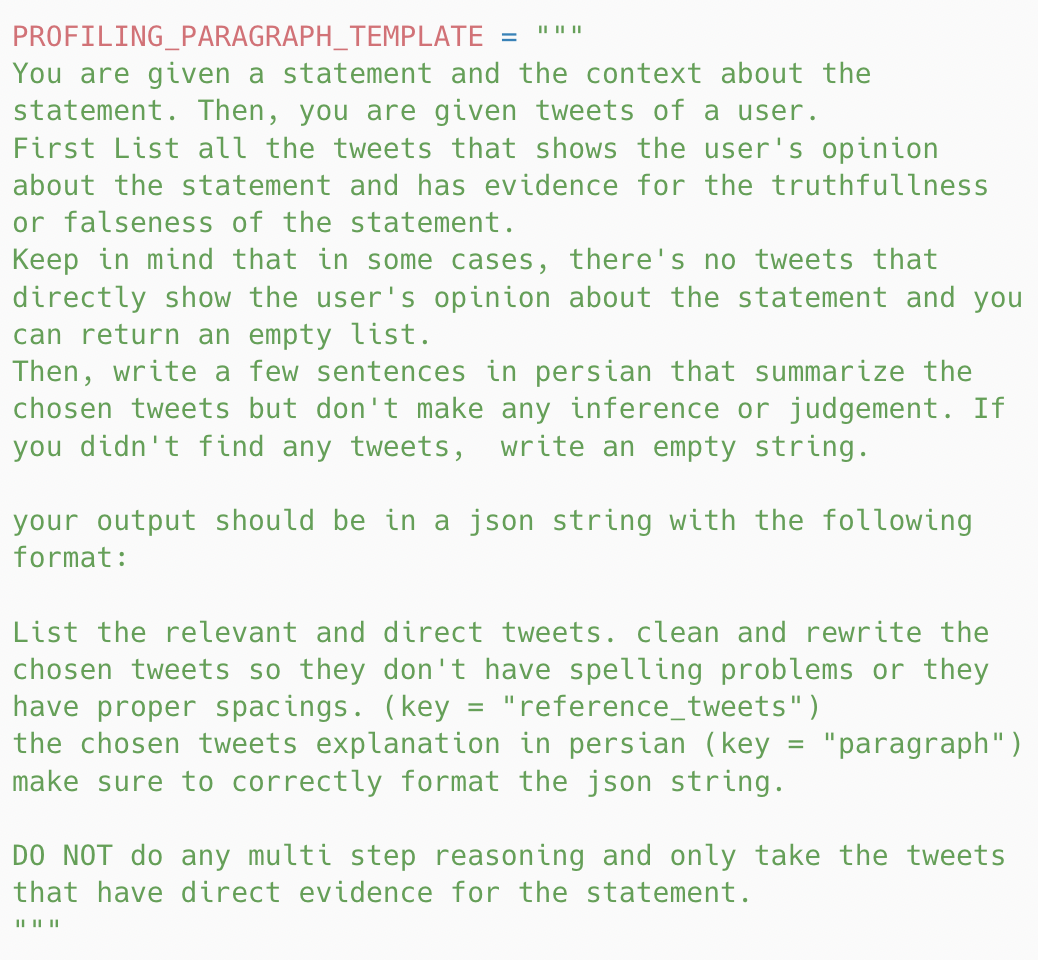}
	\caption{Prompt utilized to create our user profiles}
	\label{fig:profiling_prompt}
\end{figure}

In our profiling approach, we provide the LLM with the complete set of pooled tweets as reference context alongside each stance claim. We guide the LLM to generate factual summaries about the stance claims (Figure~\ref{fig:profiling_prompt}), explicitly instructing it to avoid reasoning or inference due to stance detection complexity and potential attention limitations with large text inputs \cite{long_context_attention} as well as LLM biases. To ensure groundedness \cite{groundedness}, we require citation of reference tweets used in the summary generation.

This yields two categories of profiles per user. Let $S = \{s_1, s_2, ..., s_n\}$ be the set of stance claims, and $T_u = \{t_1, t_2, ..., t_m\}$ be the set of tweets for user $u$. For each stance claim $s_i$, the LLM:
\begin{enumerate}
	\item Generates a factual summary $f_i$ based on relevant tweets, forming the user's \textit{Abstractive profile} $A_u = \{f_1, f_2, ..., f_n\}$
	\item Selects a set of supporting tweets $E_{u,i} \subseteq T_u$ cited in generating each summary, forming the user's \textit{Extractive profile} $E_u = \{E_{u,1}, E_{u,2}, ..., E_{u,n}\}$.
	
\end{enumerate}

\begin{figure}[!t]
	\centering
	\includegraphics[width=1\linewidth]{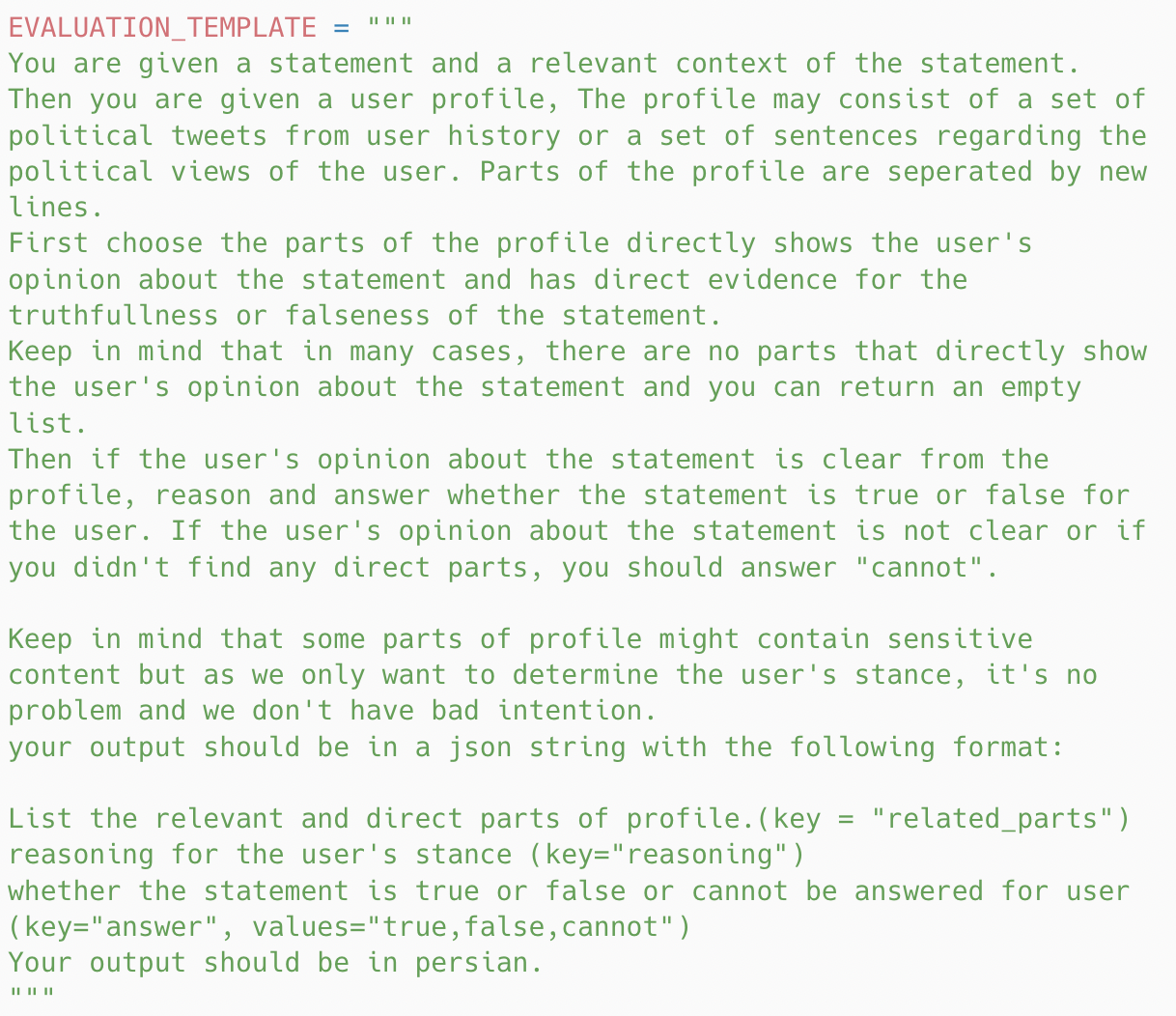}
	\caption{Prompt utilized for evaluation step}
	\label{fig:evaluation_prompt}
\end{figure}

\subsection{LLM-Based Profile Evaluation}
Our intrinsic evaluation approach is based on the intuition of information loss, measuring how well a short user profile reflects the user compared to their pooled history. We perform an open-book QA task based on the user's pooled tweet history and compare the results to the same task using just the short profiles, with the premise that better profiles will yield performance closer to using the full history.

For our selected political domain, the QA task aligns with stance detection on predefined claims. We created an evaluation dataset with 15 political stance claims for 100 users, with human annotators labeling each user-statement pair as True, False, or Cannot be answered. The annotation process involved two primary annotators with a third resolving disagreements, achieving a Cohen's kappa of 0.63. This inter-annotator agreement is satisfactory given the task's inherent subjectivity and comparable to recent Persian stance detection studies \cite{pilehvar}, with complete annotation guidelines available in Appendix~\ref{appendix:guidelines}. To evaluate profiling methods, we employ a zero-shot prompt approach using the profiles as context in a RAG setting (Figure \ref{fig:evaluation_prompt}), comparing stance detection performance against human-annotated ground truth to determine which technique most effectively preserves essential stance information.

\section{Experiments}
We categorize user profiling methods into two primary approaches:

\textbf{Extractive methods} select a diverse or representative set of tweets, preserving original content and offering privacy benefits, though they may miss deeper inferrable insights.

\textbf{Abstractive methods}, typically LLM-based, generate summaries by reasoning over tweet content, creating more compact and potentially insightful profiles. However, they raise privacy concerns through unintended pattern inference and risk introducing LLM biases that may misrepresent users' viewpoints or disproportionately emphasize certain stances.

Our study explores both approaches through comprehensive evaluation of diverse profiling techniques. Since no prior work specifically utilizes LLMs for user profiling on Twitter (X), we included methods from adjacent domains such as recommender systems and LLM personalization. Our baseline methods include random selection, BM25 retrieval \cite{BM25}, and semantic retrieval. For abstractive approaches, we evaluated the Amazon LLM personalization method and its variation with a dense retriever \cite{amazon_paper}. In extractive profiling, we implemented the Semantic Auto-Encoder (SemAE) \cite{semae}. We benchmarked our proposed profiling methods against these techniques to provide thorough comparison.

To evaluate profile quality, we employ a \textit{zero-shot stance detection} task (prompt shown in Figure~\ref{fig:evaluation_prompt}), using performance as a measure of profile richness. Ground truth labels are derived by human annotators from each user's complete pooled tweet history, providing a reliable standard for assessing how effectively each profiling method captures essential user characteristics.

\begin{figure*}[ht]
	\centering
	\includegraphics[width=1\linewidth]{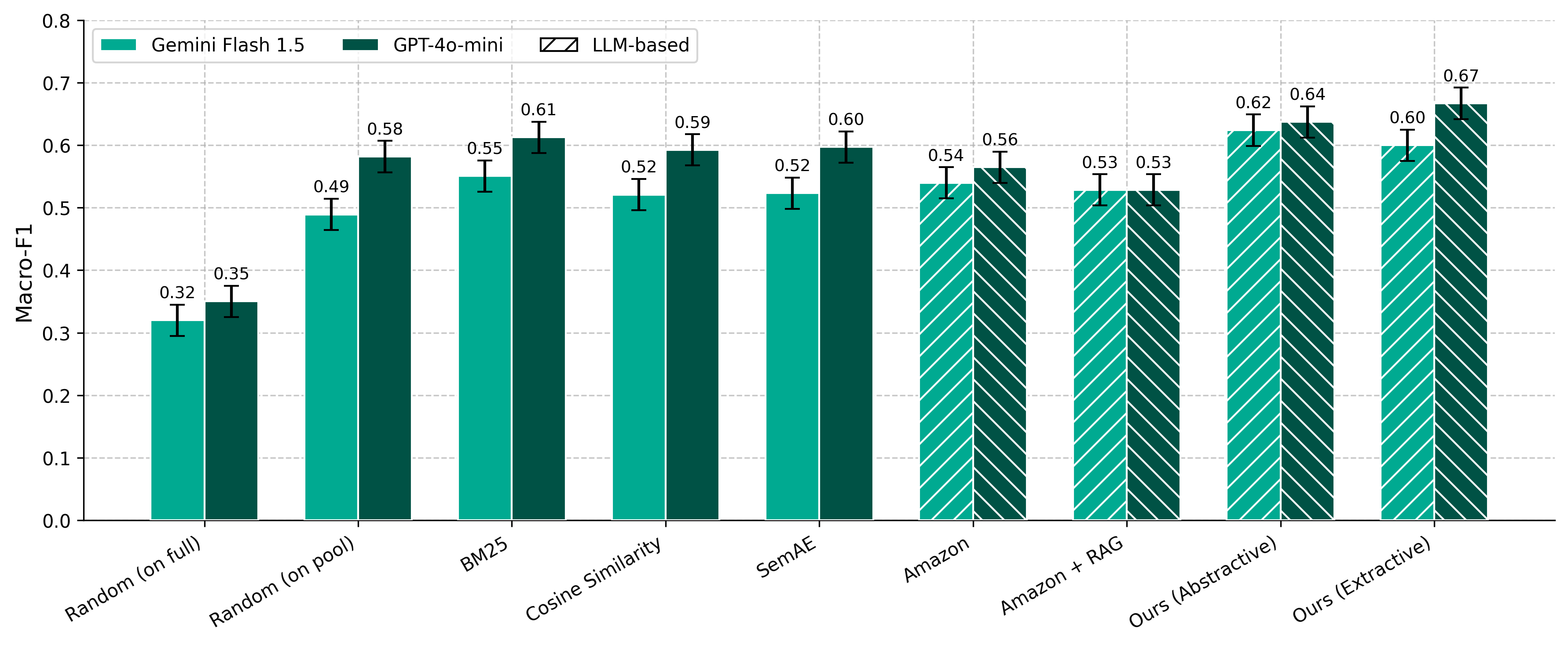}
	\caption{Macro-F1 scores of different user profiling methods evaluated using Gemini Flash 1.5 and GPT 4o mini models. Error bars represent 95\% confidence intervals. Our proposed extractive and abstractive approaches consistently outperform baseline and state-of-the-art methods across both evaluation models.}	\label{fig:experiment}
\end{figure*}

\paragraph{Standardized Profile Size} 
To ensure fairness across all methods, we standardized the profile size. Since our profiling method generates a few sentences for each of the 15 statements, we limited other selection methods to retrieve the top-1 result for each statement, resulting in 15 tweets per user. The final user profiles consist of the retrieved or generated insights for all statements combined. Importantly, these profiles remain consistent for evaluation across all user-statement pairs. For example, for user \(U\), the same profile is used for all statements from \(s_1\) to \(s_{15}\), ensuring consistency in the evaluation process.

Our stance detection task involves 1,500 user-statement pairs, covering 15 statements across 100 users. These 15 statements encompass a wide and diverse set of concepts in the Persian political domain as shown in Table~\ref{table:translated-statements}, allowing us to comprehensively evaluate how effectively each profiling method captures users' political viewpoints. Each method's processes and details are outlined below.

\subsection{Baselines}
For baseline methods, we employed several approaches to evaluate our profiling methodology:

\begin{itemize}
	\item \textbf{Random Selection:} Tweets from each user's data were randomly chosen. This baseline was implemented in two variations: one operating on our pooled tweets (maximum 80 tweets per user), and another using the complete user tweet history. 
	\item \textbf{BM25 Retriever \cite{BM25}:} BM25 is a bag-of-words information retrieval model that ranks tweets for each statement by text overlap and term frequency.
	\item \textbf{Semantic Retrieval:} This method measures cosine similarity between semantic embeddings of tweets and stance claim to retrieve tweets relevant to it.
\end{itemize}

\noindent
Details of the implementation and parameters for these baselines are provided in the Appendix~\ref{appendix:implementation-details}.

\subsection{Abstractive Profiling }
The Amazon approach focuses on generating abstractive summaries using an LLM for user profiling. Two variations were evaluated:
\begin{itemize}
	\item \textbf{Amazon Summarization:} This method generates an abstractive summary over the entire user history as a single output.
	\item \textbf{Amazon with RAG:} In this variation, summaries are generated for each of the 15 stance claims individually by retrieving relevant tweets and then aggregating the summarizations. 
\end{itemize}
The prompts used for these tasks are detailed in Appendix~\ref{appendix:prompts}. 

\subsection{Extractive Profiling}
The Semantic Auto-Encoder (SemAE) \cite{semae} uses aspect-based summarization to extract tweets relevant to stance-specific topics:
\begin{itemize}
	\item Related tweets are extracted by matching keywords associated with each stance.
	\item The method computes the mean semantic embedding of the given stance and selects tweets nearest to this embedding as the user’s profile.
\end{itemize}
The specific keywords adapted for each of the stances are detailed in Appendix~\ref{appendix:implementation-details}. 

\section{Results and Discussion}

The results in Fig.~\ref{fig:experiment} illustrate the performance of various user profiling methods for political stance detection, with bar charts displaying macro F1 scores and their 95\% confidence intervals. Both variations of our approach outperform existing baseline and state-of-the-art methods, with the extractive approach showing particularly strong results.

\subsection{Extractive vs. Abstractive Profiling}

\subsubsection{Impact of Cultural and Political Bias.}
Our error analysis reveals that LLMs exhibit inherent biases stemming from their Western-centric training data. This particularly affects the abstractive approach when handling region-specific political concepts such as "reformist," which carries different connotations in the Persian political landscape. These cultural nuances are better preserved in the extractive method, where the model selects actual user statements rather than generating summaries that might inadvertently incorporate Western perspectives on Middle Eastern conflicts.

\subsubsection{Attention and Context Length Limitations.}
The performance gap between approaches can also be attributed to LLMs' limitations in handling complex multi-step reasoning with long contexts. When implementing the abstractive approach using the prompt in Fig.~\ref{fig:profiling_prompt}, GPT-4o-mini must simultaneously process numerous tweets, identify patterns, and generate coherent paragraphs—triggering challenges in the attention mechanism \cite{long_context_attention}. The extractive method, on the other hand, simply selects the most relevant tweets towards stance claims from the user history.

\subsection{Performance Comparison Across Methods}
\subsubsection{Statistical Significance of Results.}
Our extractive approach achieved a macro F1 score of 0.6668 with GPT-4o as evaluator, significantly outperforming all other methods with 100\% statistical significance with McNemar's test. When evaluated using Flash, both our methods demonstrated statistically significant improvements in 85.7\% of comparisons. Detailed information of tests using McNemar's test are provided in Appendix~\ref{appendix:statistics}.
\subsubsection{Performance of Baseline Methods.}
Random selection from our curated pool performed relatively well due to our evaluation design, which allowed selection from a high-quality pool of approximately 80 tweets. This increases the likelihood of including stance-relevant content, particularly effective because users' political stances tend to remain consistent over time. The Amazon approach underperformed largely because its minimal summarization prompt generated overly brief outputs (typically just one or two paragraphs), while other methods incorporated up to 15 tweets for more comprehensive and nuanced profiles. This underperformance highlights how a carefully crafted prompt, such as the one used in our experiments, significantly impacts the quality of profile generation. Interestingly, sparse retrieval (BM25) outperformed dense retrieval methods in our tests. Our error analysis revealed this was primarily due to the importance of specific keywords in stance detection tasks, especially when evaluating positions regarding political figures like presidents—where exact name matching provided by sparse retrieval offers a distinct advantage over semantic matching, even when using state-of-the-art multilingual embedding models like BGE-M3~\cite{bge-m3}. Aspect Summarization in SemAE achieved comparable results to retrieval-based methods by extracting related tweets using keywords and computing mean embeddings to select the most representative tweets, demonstrating the effectiveness of strategic tweet selection in building user profiles.
\subsection{Challenges in Stance Detection.}
Stance detection in Persian remains challenging, as demonstrated by \cite{pilehvar}, where even GPT-4o achieves an accuracy of only 74.8\% on Persian stance detection tasks. The subjective nature of this task is evidenced by the Cohen's kappa score of 0.64 between our annotators. However, our work focuses primarily on comparing different user profiling methods, and since LLM-generated evaluation limitations are consistent across all methods, our comparisons provide a robust baseline.

\section{Conclusion}

In this paper, we introduced a novel approach to user profiling by analyzing over six million tweets using a domain-specific knowledge base combined with large language models. Our method overcomes traditional challenges such as reliance on large labeled datasets while improving interpretability and adaptability. Through semi-supervised filtering and LLM reasoning capabilities, we generated concise, interpretable user profiles that significantly outperform all baselines and state-of-the-art techniques in our evaluations.

Our observations show that abstractive profiles are susceptible to inherent LLM biases, which we intend to address in future work through knowledge integration and comprehensive bias evaluation frameworks. To maintain focus in this initial investigation, we concentrated on establishing our core methodology; future work will expand to evaluate these profiles across various downstream tasks and domains, integrate targeted domain-specific knowledge, and develop specialized embedding models for improved filtering.

As LLMs increasingly perform reasoning tasks across social networks, our work contributes by offering a method to distill expansive user histories into computationally efficient profiles that LLMs can effectively utilize, opening new possibilities for personalized AI applications and computational social science research.

\section*{Acknowledgments}
This research was in part supported by a grant from the School of Computer Science,
Institute for Research in Fundamental Sciences, IPM, Iran (No. CS1403-4-05).

\bibliography{main}

\appendix

\section{Appendix}

\subsection{Data Details}
\label{appendix:data-details}

In this section we have the extra full version of relevant data used and generated during the research.

\begin{table}[ht!]
	\centering
	\setlength{\tabcolsep}{5pt}  
	\begin{tabular}{p{0.95\columnwidth}}  
		\hline
		\textbf{Political Stance Claims} \\
		\hline
		Positive opinion about current government's performance \\
		Support for Iran's foreign policies \\
		Support for JCPOA (Iran Nuclear Agreement) \\
		Positive opinion about mandatory hijab \\
		Positive opinion about political prisoners' situation \\
		Support for electoral participation \\
		Positive opinion about foreign media outlets \\
		Support for reducing Afghan immigration \\
		\hline
	\end{tabular}
	\caption{Example Political Stance Claims}
	\label{table:translated-statements}
\end{table}

\subsection{Prompts}

\label{appendix:prompts}

In this section we have the prompts used in different stages of the pipeline.

\begin{figure}[ht]
	\centering
	\includegraphics[width=\columnwidth]{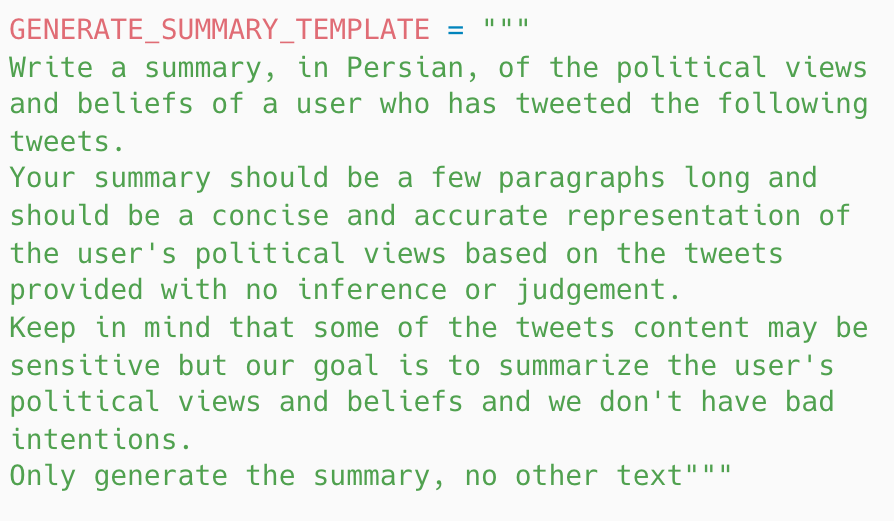}
	\caption{The summarization prompt used in the Amazon approach to summarize content}
	\label{fig:amazon_summarization_prompt}
\end{figure}

\begin{figure}[ht]
	\centering
	\includegraphics[width=\columnwidth]{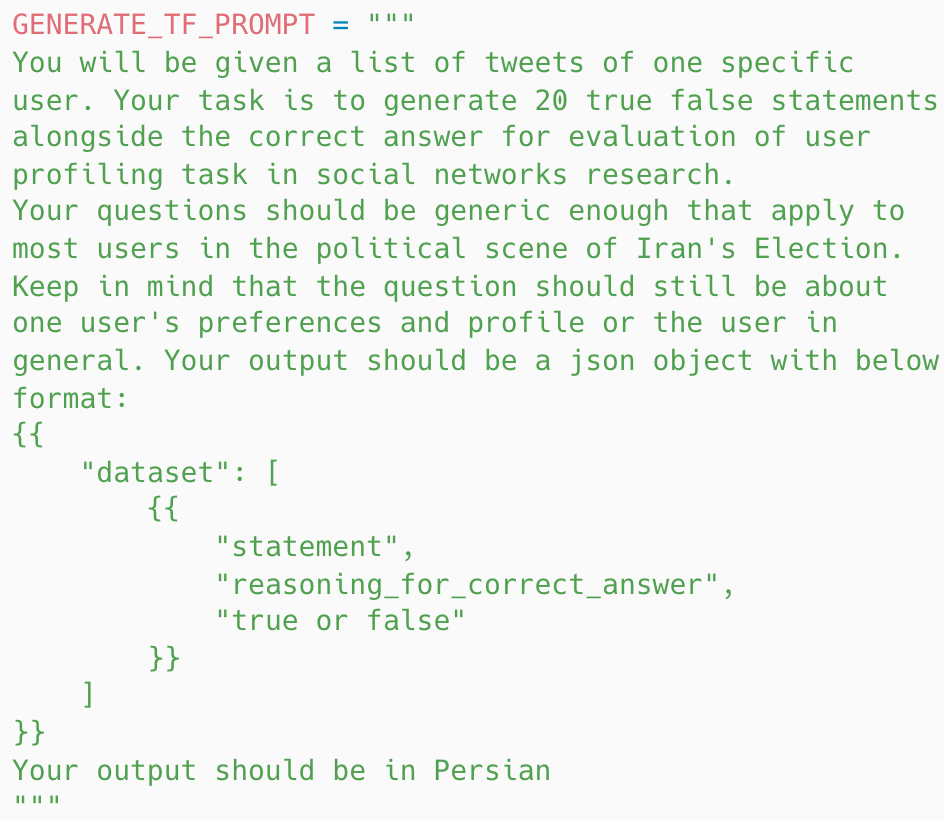}
	\caption{The prompt used during the statement generation stage for defining domain statements}
	\label{fig:statement_generation_prompt}
\end{figure}

\subsection{Annotation Guidelines}
\label{appendix:guidelines}

Annotators are provided with a stance claim and a pool of up to 100 tweets from a user's history, curated to be comprehendable by human reviewers. For each user-statement pair, annotators must classify the statement as either True, False, or Cannot Be Answered based solely on the provided tweet context. 

\textbf{Annotation Rules:}
\begin{itemize}
	\item \textbf{True:} Direct evidence exists in the user's history supporting the statement (e.g., user explicitly agrees with government's economic decisions)
	\item \textbf{False:} Direct evidence exists in the user's history contradicting the statement (e.g., user explicitly opposes reformist political movement)
	\item \textbf{Cannot Be Answered:} No direct evidence exists in the user's history to determine the truthfulness of the statement
\end{itemize}

\textbf{Important Constraints:}
\begin{itemize}
	\item Multi-step reasoning and inference are prohibited
	\item Annotations must be based solely on explicit evidence in the provided tweets
	\item Maximum 300 user-statement pairs can be annotated per day to maintain quality
\end{itemize}

\subsection*{Statistical Significance Tests}
\label{appendix:statistics}

In this section, we provide the results of statistical significance tests using McNemar's test. The p-values and significance results for comparisons between our methods and baselines are summarized in Table~\ref{table:statistics}. Significance (\textit{p < 0.05}) is marked as \textbf{Y} (Yes) or \textbf{N} (No).

\begin{table}[!h]
	\centering
	\caption{Statistical Significance Tests (McNemar's Test) for both Flash and GPT evaluations. Significance (\textit{Y}/\textit{N}) indicates whether the result was statistically significant (\textit{p < 0.05}).}
	\label{table:statistics}
	\begin{tabular}{|l|c|c|}
		\hline
		\textbf{Against Method}      & \textbf{Flash (p, Sig.)} & \textbf{GPT (p, Sig.)} \\ \hline
		\multicolumn{3}{|c|}{\textbf{Ours (Abstractive)}} \\ \hline
		Random                       & 0.0000, Y               & 0.0878, N              \\ \hline
		BM25                         & 0.0001, Y               & 0.5333, N              \\ \hline
		RAG                          & 0.0000, Y               & 0.2778, N              \\ \hline
		SemAE                        & 0.0000, Y               & 0.4137, N              \\ \hline
		Amazon                       & 0.0000, Y               & 0.0000, Y              \\ \hline
		Amazon+RAG                   & 0.0000, Y               & 0.0000, Y              \\ \hline
		Ours (Extractive)            & 0.1981, N               & 0.0002, Y              \\ \hline
		\multicolumn{3}{|c|}{\textbf{Ours (Extractive)}} \\ \hline
		Random                       & 0.0000, Y               & 0.0000, Y              \\ \hline
		BM25                         & 0.0016, Y               & 0.0007, Y              \\ \hline
		RAG                          & 0.0001, Y               & 0.0001, Y              \\ \hline
		SemAE                        & 0.0002, Y               & 0.0003, Y              \\ \hline
		Amazon                       & 0.0003, Y               & 0.0000, Y              \\ \hline
		Amazon+RAG                   & 0.0001, Y               & 0.0000, Y              \\ \hline
		Ours (Abstractive)           & 0.1981, N               & 0.0002, Y              \\ \hline
	\end{tabular}
\end{table}

\subsection{Implementation Details}
\label{appendix:implementation-details}

In this section we go over the experiment implementation details for different steps of user profiling and benchmarking

\begin{algorithm}[!ht]
	\caption{Iterative Tweet Selection}
	\footnotesize
	\label{alg:tweet_selection}
	\textbf{Input}: $group$ (tweets with vector embeddings), $n\_select$ (number of tweets to select), $initial\_threshold$ (cosine similarity threshold) \\
	\textbf{Output}: Selected tweets ($selected\_tweets$)
	\begin{algorithmic}[1]
		\STATE Compute mean embedding from tweet vectors.
		\STATE Calculate cosine similarity of each tweet to the mean embedding.
		\STATE Sort tweets by similarity (descending).
		\STATE Initialize 
		
		$selected\_indices \gets \emptyset$, 
		
		$unchecked\_indices \gets \{0, 1, \dots, |group|-1\}$, 
		
		$threshold \gets initial\_threshold$,
		
		$iteration \gets 0$.
		\WHILE{$|selected\_indices| < n\_select$ and $|unchecked\_indices| > 0$}
		\STATE Select the tweet with the highest similarity ($current\_index \gets \text{unchecked\_indices}[0]$).
		\STATE Add $current\_index$ to $selected\_indices$ and remove it from $unchecked\_indices$.
		\STATE Compute cosine similarity of $current\_index$ to remaining tweets in $unchecked\_indices$.
		\STATE Remove indices from $unchecked\_indices$ with similarity $\geq threshold$.
		\STATE Update $threshold \gets \text{logarithmic\_decay}(initial\_threshold, iteration)$.
		\STATE Increment $iteration \gets iteration + 1$.
		\ENDWHILE
		\IF{$|selected\_indices| < n\_select$}
		\STATE Add tweets from $unchecked\_indices$ to meet $n\_select$.
		\ENDIF
		\STATE \textbf{return} tweets corresponding to $selected\_indices$
	\end{algorithmic}
\end{algorithm}
\begin{algorithm}[!ht]
	\caption{Stratified Sampling of Tweets based on Clustering}
	\footnotesize
	\label{alg:tweet_clustering}
	\textbf{Input}: $group$ (tweets with vector embeddings), $n\_select$ (number of tweets to sample) \\
	\textbf{Output}: Selected tweets ($selected\_tweets$)
	\begin{algorithmic}[1]
		\STATE Compute embeddings for all tweets in $group$.
		\STATE Use the Elbow method to find the optimal $k$ for $K$-Means clustering:
		\begin{itemize}
			\item For each $k$ in a range of possible cluster counts:
			\begin{itemize}
				\item Apply $K$-Means clustering with $k$ clusters.
				\item Compute within-cluster sum of squares (WCSS).
			\end{itemize}
			\item Plot WCSS vs. $k$ and choose the "elbow point" where decrease flattens.
		\end{itemize}
		\STATE Apply $K$-Means clustering using the optimal $k$ found in the previous step.
		\STATE Assign each tweet to one of the $k$ clusters.
		\STATE Compute the sampling quota for each cluster: \\ 
		$quota_i = \frac{|C_i|}{|group|} \times n\_select$, where $|C_i|$ is the size of cluster $i$. 
		\STATE For each cluster:
		\begin{itemize}
			\item Randomly sample $quota_i$ tweets without replacement.
		\end{itemize}
		\STATE Combine sampled tweets from all clusters.
		\STATE \textbf{return} $selected\_tweets$
	\end{algorithmic}
\end{algorithm}

\begin{table*}[ht!]
	\centering
	\setlength{\tabcolsep}{8pt} 
	\begin{tabular}{|p{6cm}|p{9.5cm}|}
		\hline
		\textbf{Statement} & \textbf{Keywords} \\ 
		\hline

		Positive opinion about the current government’s performance &
		Government, Performance, Support, Criticism, Cabinet, Economy, Minister, Policy, Problem, Progress, Livelihood, Dissatisfaction, Boost \\ 
		\hline
		
		Supports the JCPOA (Iran Nuclear Agreement) &
		JCPOA, Agreement, Rouhani, Obama, Trump, Suspension, Sanction, Diplomacy, Negotiation, Peace, Return, Commitment, Interests \\ 
		\hline
		
		Positive opinion about mandatory hijab &
		Hijab, Mandatory, Patrol, Guidance, Control, Modesty, Law, Women, Clothing, Ethics, Values, Opposition, Preservation, Culture \\ 
		\hline
		
		Positive opinion about the situation of political prisoners &
		Prisoner, Political, Evin, Positive, Convicted, Rights, Freedom, Oppressed, Imprisonment, Crime, Court, System, Law, Prison \\ 
		\hline
		
		People should participate in elections &
		Election, Participation, People, Vote, Right, Duty, Democracy, Choice, Freedom, Decision, Responsibility, Politics, Voter, Voting \\ 
		\hline
		
		Positive opinion about foreign media outlets &
		Media, Foreign, Satellite, Iran, BBC, Positive, Channel, News, Publication, Information, Lie, Opposition, Advertisement, Truth \\ 
		\hline

		Supports reducing the number of Afghan migrants in Iran &
		Migrant, Afghan, Reduction, Illegal, Number, Nationality, Iranian, Return, Problem, Border, Control, Citizen, Criticism, Migration \\ 
		\hline

	\end{tabular}
	\caption{Domain-Adapted Translated Example Keywords Used with the SemAE Approach}
	\label{table:semae-keywords}
\end{table*}

\subsubsection{Information Retrieval Systems.}
For the BM25 retrieval implementation, we utilized the Tantivy Python package. The cosine similarity calculations were performed using LanceDB with nprobes set to 300. For embedding calculations used with LanceDB, we employed the sentence-transformers version of BGE-M3.

\subsubsection{Language Model Configuration.}
The large language model calls were made through OpenRouter.ai with the following parameters:
\begin{itemize}
	\item Temperature: $1\times10^{-100}$
	\item Top-p: 1
	\item Top-k: 1
\end{itemize}
The implementation was built using LangChain framework for software architecture and prompt templating.

\subsubsection{BERT Fine-tuning Parameters.}
The Tooka-BERT model was fine-tuned using PyTorch with the following configuration:
\begin{itemize}
	\item Training batch size: 128 (per device)
	\item Evaluation batch size: 128 (per device)
	\item Gradient accumulation steps: 8
	\item Number of epochs: 15
	\item Learning rate: $7\times10^{-4}$
	\item Weight decay: 0.01
	\item Warmup steps: 500
\end{itemize}

\subsubsection{Training Configuration.}
The training process included:
\begin{itemize}
	\item Evaluation strategy: Steps-based with evaluation every 10 steps
	\item Model saving: Every 100 steps with total limit of 1
	\item Early stopping: Patience of 6 steps with 0.01 improvement threshold
	\item Metrics: Accuracy-based model selection
	\item Logging: TensorBoard integration with 2-step logging frequency
\end{itemize}
\subsection{Adaptation to other Domains}
\label{appendix:adaptation}
To adapt our profiling approach to a new domain, the primary requirement is creating an initial knowledge base if one does not already exist. In this section, we demonstrate this adaptation process using Wikidata knowledge graph in the European football domain. This implementation serves as an illustrative example rather than a comprehensive solution. We begin by selecting relevant node types and edge types from Wikidata as depicted in Table~\ref{tab:football}, then traverse the graph to our preferred depth to collect a more comprehensive set of nodes. For each identified node, we retrieve the corresponding Wikipedia pages, which form our initial knowledge base. Once this domain-specific knowledge base is established, our proposed method can be applied as previously described.
\begin{table}[H]
	\centering
	\setlength{\tabcolsep}{6pt}  
	\begin{tabular}{ll}
		\hline
		\textbf{Nodes (Competitions/Leagues)} & \textbf{Wikidata ID} \\
		\hline
		\multicolumn{2}{l}{\textbf{Major Competitions}} \\
		\hline
		Champions League & Q18756 \\
		UEFA Europa League & Q18760 \\
		European Championship & Q260858 \\
		\hline
		\multicolumn{2}{l}{\textbf{Top Leagues}} \\
		\hline
		Premier League & Q9448 \\
		La Liga & Q324867 \\
		Bundesliga & Q82595 \\
		Serie A & Q15804 \\
		Ligue 1 & Q13394 \\
		\hline
		\multicolumn{2}{l}{\textbf{Entity Types}} \\
		\hline
		Football Club & Q476028 \\
		Football Player & Q937857 \\
		Football Manager & Q628099 \\
		\hline
		\multicolumn{2}{l}{\textbf{Edges (Relationships)}} \\
		\hline
		P54 & Member of sports team \\
		P1344 & Participant of \\
		P106 & Occupation \\
		P286 & Head coach \\
		P413 & Position played \\
		\hline
	\end{tabular}
	\caption{Node and Edge Types for European Football Domain}
	\label{tab:football}
\end{table}

\end{document}